# Mapping between Spin-Glass Three-Dimensional (3D) Ising Model and Boolean Satisfiability Problem

Zhidong Zhang

Shenyang National Laboratory for Materials Science, Institute of Metal Research, Chinese Academy of Sciences, 72 Wenhua Road, Shenyang, 110016, P.R. China

**Abstract:**

The common feature for a nontrivial hard problem is the existence of nontrivial topological structures, non-planarity graphs, nonlocalit**ies**, or long-range spin entanglements in a model system with randomness. For instance, the Boolean satisfiability (K-SAT) problems for K ≥ 3 $M_{SAT}^{K \geq 3}$ are nontrivial, due to the existence of non-planarity graphs, nonlocalit**ies**, and the randomness. In this work, the relation between a spin-glass three-dimensional (3D) Ising model $M_{SGI}^{3D}$ with the lattice size $N = mnl$ and the K-SAT problems is investigated in detail. With the Clifford algebra representation, it is easy to reveal the existence of the long-range entanglements between Ising spins in the spin-glass 3D Ising lattice. The internal factors in the transfer matrices of the spin-glass 3D Ising model lead to the nontrivial topological structures and the nonlocalit**ies.** At first, we prove that the absolute minimum core (AMC) model $M_{AMC}^{3D}$ exists in the spin-glass 3D Ising model, which is defined as a spin-glass 2D Ising model interacting with its nearest neighboring plane. Any algorithms, which use any approximations and/or break the long-range spin entanglements of the AMC model, cannot result in the exact solution of the spin-glass 3D Ising model. Second, we prove that the dual transformation between the spin-glass 3D Ising model and the spin-glass 3D $Z_2$ lattice gauge model shows that it can be mapped to a K-SAT problem for K ≥ 4 also in the consideration of random interactions and frustrations. Third, we prove that the AMC model is equivalent to the K-SAT problem for K = 3. Because the lower bound of the computational complexity of the spin-glass 3D Ising model $C_L(M_{SGI}^{3D})$ is the computational complexity by brute force search of the AMC model $C^U(M_{AMC}^{3D})$, the lower bound of the computational complexity of the K-SAT problem for K ≥ 4 $C_L(M_{SAT}^{K \geq 4})$ is the computational complexity by brute force search of the K-SAT problem for K = 3 $C^U(M_{SAT}^{K=3})$. Namely, $C_L(M_{SAT}^{K \geq 4}) = C_L(M_{SGI}^{3D}) \geq C^U(M_{AMC}^{3D}) = C^U(M_{SAT}^{K=3})$. All of them are in subexponential and superpolynomial. Therefore, the computational complexity of the K-SAT problem for K ≥ 4 cannot be reduced to that of the K-SAT problem for K < 3.

The corresponding author: Z.D. Zhang,

Tel: 86-24-23971859,

Fax: 86-24-23891320,

e-mail address: zdzhang@imr.ac.cn

## 1. Introduction

In recent years, there has been great progress in computer science, specially, in machine learning, artificial intelligence, data mining, and so on. The technical advances in these fields change our daily life, while they also benefit our better understanding on the fundamental structures of mathematics. Meanwhile, there is a trend to strengthen the discipline among mathematics, physics, and computer science to solve hard problems in science. Indeed, to solve a problem in physics, one may need to understand in depth the mathematical structures of a physical system, which may involve knowledge in algebra, topology and geometry. By contrast, a hard problem in mathematics and computer science may be related to a physical system, and to solve it, one may seek the guidance of physical significances. Moreover, successfully solving a hard problem in either physics or mathematics/computer science may provide a new forum for dialogues between mathematicians, physicists, as well as computer scientists.

The exact solution of a ferromagnetic three-dimensional (3D) Ising model in a zero external magnetic field is a well-known problem in physics, standing for almost 100 years [1–6]. To explicitly solve it, we need to understand well the mathematical structures with algebraic, topological, and geometric aspects [1–8], for instance, Jordan algebra [9–12]. The procedures for exactly solving it involve employing several algebras (Lie algebra, Jordan algebra, Clifford algebra, quaternion algebra, etc.), accounting for the contribution of the nontrivial topological structures to the physical properties, and generating the geometric phases on eigenvectors and eigenvalues of the ferromagnetic 3D Ising system [3,5–8,13]. The 3D Ising model is related closely with combinatorics, graph theory, and statistical learning networks. A famous example in machine learning and artificial intelligence is Alphago that defeated the world champion of the game of Go [14], with technical advances in deep learning and artificial neural networks. The game of Go deals with computing the total state of a system with three possible states (black, while, empty) at each site of a 19 × 19 lattice, which actually corresponds to a two-dimensional (2D) q = 3 Potts model with a spin of three possible states (spin up, spin down, empty; or with values +1, −1, 0) at each site. Note that the Ising model is a q = 2 Potts model with spin up and spin down states (with values +1/2, −1/2). The computational complexity of the spin-glass 3D Ising model is much more complicated than that of the game of Go because the lattice is 3D, resulting in the nontrivial topological structures and the nonlocality (due to the long-range entanglement between Ising spins, a pure quantum mechanics effect), and the lattice size is $N = mnl$, where $m$, $n$, $l$ are the numbers of lattice points along three crystallographic directions, and in the thermodynamic limit $m \to \infty$, $n \to \infty$, $l \to \infty$, $N \to \infty$ [15].

In computational complexity theory, NP is an abbreviation for non-deterministic polynomial time, which is defined as the set of decision problems that can be solved in polynomial time on a non-deterministic Turing machine. A P-problem can be solved in polynomial time by a deterministic Turing machine. In NP, the set of all decision problems whose solutions can be verified in polynomial time are cataloged to NP-complete problems (Cook–Levin theorem [16,17]). A problem p in NP is NP-complete if every other problem in NP can be transformed into p in polynomial time. The most notable characteristic of NP-complete problems is that no fast solution to them is known. The K-satisfiability (K-SAT) problem for K ≥ 3 is a central problem in combinatorial optimization, being the first problem to be shown NP-complete. The K-SAT problem deals with an ensemble of N Boolean variables, submitted to M constraints. Each constraint is expressed in the form of an OR function of K variables (or their negations) in the ensemble, and the problem is to check whether there exists one configuration

(among $2^N$ possible ones) of the variables, which satisfies all constraints. An efficient algorithm for solving the K-SAT problem in its worst-case instances will immediately lead to other algorithms for solving efficiently thousands of different hard combinatorial problems. Among thousands of NP-complete problems [18–31], we mention several well-known ones as follows: Hamiltonian path problem, travelling salesman problem, Knapsack problem, subset sum problem, vertex cover problem, subgraph isomorphism problem, independent set problem, graph coloring problem, dominating set problem, protein folding problem, maximum edge biclique problem, etc. At present, all known algorithms for NP-complete problems require time that is super-polynomial in the input size, and it is unknown whether there are any faster algorithms. In the previous work [32–34], the spin-glass 2D Ising model was proven to be a P-problem, whereas the spin-glass 3D Ising model was proven to be a NP-complete problem. There have been some reports [35–41] on the relation between the spin-glass Ising models and K-SAT problems.

Although the NP-complete problem is an important problem in computer sciences, it consists of thousands of problems in different fields, such as mathematics, physics, chemistry, biology, and so on. It is thought that any advances in these fields for anyone of these problems may shed a certain light on solving this NP-complete problem. The present author has been working on the 3D Ising models for tens of years and figures out their mathematical structures [3–8,15,42], which are quite helpful for understanding the spin-glass 3D Ising model. The aim of this work is to investigate the mapping between the spin-glass 3D Ising model and the K-SAT problems, with emphasis of the Clifford algebra representation and the dual transformation to reveal the nontrivial topological structures, the non-planarity graphs, the nonlocalities, or the long-range spin entanglements in these two systems. It is important to use the Clifford algebra representation to reveal the long-range entanglement (as well as the nonlocality) between Ising spins in the spin-glass 3D Ising lattice, which, together with the existence of the randomness, causes the non-triviality of the problem. It is found that the spin-glass 3D Ising model can be mapped by a dual transformation to a K-SAT problem for K ≥ 4. If we focus on an absolute minimum core (AMC) model of the spin-glass 3D Ising model, which is defined as a spin-glass 2D Ising model interacting with its nearest neighboring plane, the lower bound of the computational complexity of the spin-glass 3D Ising model (as well as the K-SAT problems for K ≥ 4) is reduced to the computational complexity by brute force search of the AMC model (as well as a K-SAT problem for K = 3).

The paper is organized as follows: In Section 2, the Clifford algebra structure and nonlocality of the spin-glass 3D Ising model are investigated. In Section 3, a mapping between the spin-glass 3D Ising model and the Boolean satisfiability for K ≥ 4 is established. In Section 4, we figure out the mapping between the AMC model and the K-SAT problem for K = 3. Section 5 is for conclusions.

## 2. Nonlocality of the Spin-Glass 3D Ising Model

In nature, there exist different magnetic materials with, for instance, ferromagnetic, antiferromagnetic, ferrimagnetic, paramagnetic, spin glass, and even spin liquid phases. The formation of these magnetic phases and the phase transitions between them are governed by competition between terms of various energies, including the exchange interactions, the crystalline anisotropy energy, the Zeeman energy (caused by an external magnetic field), and thermal activity. The existence of these magnetic states is controlled by the minimum of the total free energy of the system. In a ferromagnet, the ferromagnetic ordered state emerges at the critical point, which is spontaneously magnetized with unit vector denoting direction of saturation magnetization. The spins in a ferromagnet all align in the same direction in its ground state, while in an antiferromagnet, the neighboring spins are antiparallelly aligned in the ground state. In an Ising magnet [1–4], the spontaneous magnetization points to the z axis according to the usual definition. Above Curie (or Néel) temperature, the parametric phase appears as a disorder state of spin alignments. A spin glass is a disordered magnet [15,43–47], where the spins are aligned in an irregular pattern. All the spins in a spin glass are frozen in a disorder ground state, aligning randomly to different directions (in an Ising case, +z and -z directions). In a certain sense, the spin-glass state is an ordered state with disorder orientations of spins. The difference between spin glass and parametric phases is that, in the paramagnetic state, the spins align disorderly in space and with the time evolution; in the spin glass, the spins align disorderly in space but may remain ordered (and/or unchanged) with the time evolution (associated to the onset of the spontaneous replica symmetry breaking). Furthermore, frustration, non-ergodic behavior, and even nontrivial topological effect (in 3D) may occur in the spin glass systems [15].

**Definition 1.** *Let $M_A^D$ be a physical model where the upper script fixes the dimension, and the lower indices indicate the character of the model.*

**Definition 2.** *Let $C(M_A^D)$ be the computational complexity of the model $M_A^D$.*

**Definition 3.** *Let $C^U(M_A^D)$ be the upper bound of the computational complexity of $M_A^D$. The upper bound for a model is equal to the computational complexity as computed by brute force search.*

**Definition 4.** *Let $C_L(M_A^D)$ be the lower bound of the computational complexity of $M_A^D$.*

**Theorem 1.** *The long-range entanglement between Ising spins exists in the spin-glass 3D Ising lattice, which is represented by the exponential factors of $s'_j s'_{j+mn}$ in the transfer matrices.*

**Proof of Theorem 1.** The Hamiltonian of a spin-glass 3D Ising (Edwards–Anderson) model $M_{SGI}^{3D}$ is written as [15,44]:

$$H = -\sum_{<i,j>} J_{ij} S_i S_j \qquad (1)$$

where Ising spins with $S = 1/2$ are arranged on a 3D lattice with the lattice size $N = mnl$. The numbers ($m$, $n$, $l$) denote lattice points along three crystallographic directions. We consider only the nearest neighboring interactions $J_{ij}$ with different signs (for ferromagnetic or antiferromagnetic ones), which are randomly distributed and can be set to be different. We shall use $\tilde{J}$, $\tilde{J}'$, and $\tilde{J}''$, being a probability distribution, to represent the randomly distributed interactions along the three crystallographic directions, respectively. As usual, the probability of finding the spin-gass 3D Ising lattice in a given configuration and a fixed replica at the temperature $T$ is proportional to $exp\{-E_c/k_BT\}$, where $E_c$ is the total energy of the configuration and $k_B$ is the Boltzmann constant. The thermodynamic properties for the spin-glass 3D Ising model can be found from the partition function $Z$, after mediating $\ln\bar{Z}$ over disorder. The partition function $\bar{Z}$ for the spin-glass 3D Ising lattice in a fixed replica can be expressed as [2,3,15,48]:

$$\bar{Z} = \sum_{all\ configurations} e^{n_c \tilde{K} + n'_c \tilde{K}' + n''_c \tilde{K}''} \qquad (2)$$

Here we use $\bar{Z}$ to represent the partition function in a fixed replica, that is, the annealed average of the partition function Z. $n_c$, $n'_c$ and $n''_c$ are integers depending on the configuration of the spin lattice [2,3,48], and $\bar{Z}$ is obtained by summarizing over all configurations in a fixed replica. The variables $\widetilde{K} \equiv \tilde{J}/(k_B T)$, $\widetilde{K}' \equiv \tilde{J}'/(k_B T)$ and $\widetilde{K}'' \equiv \tilde{J}''/(k_B T)$ are introduced instead of $\tilde{J}$, $\tilde{J}'$, and $\tilde{J}''$ for describing the randomly distributed interactions. The partition function $\bar{Z}$ of the spin-glass 3D Ising lattice in a fixed replica may be written in forms of three transfer matrices in forms of direct products of matrices [3,4,7,49,50]. The following generators of Clifford algebra of the 3D Ising model are introduced:

$$\Gamma_{2k-1} = C \otimes C \otimes \ldots \otimes C \otimes s' \otimes 1 \otimes \ldots \otimes 1 \quad \text{(k-1 times } C\text{)} \tag{3}$$

$$\Gamma_{2k} = C \otimes C \otimes \ldots \otimes C \otimes (-is'') \otimes 1 \otimes \ldots \otimes 1 \quad \text{(k-1 times } C\text{)} \tag{4} \quad (3)$$

Following the Onsager–Kaufman–Zhang notation [2–4,7], we have: $s'' = \begin{bmatrix} 0 & -1 \\ 1 & 0 \end{bmatrix}$ (= $i\sigma_2$), $s' = \begin{bmatrix} 1 & 0 \\ 0 & -1 \end{bmatrix}$ (= $\sigma_3$), $C = \begin{bmatrix} 0 & 1 \\ 1 & 0 \end{bmatrix}$ (= $\sigma_1$), where $\sigma_j$ ($j = 1,2,3$) are Pauli matrices.

Because we are interested in the computational complexity of the spin-glass 3D Ising model, its partition function Z can be calculated from the average of the partition function $\bar{Z}$ for many fixed replicas. Since the computational complexity for computing the partition function Z is much more complicated than that of the annealed average $\bar{Z}$, it is enough to focus on $\bar{Z}$ for the lower bound of the computational complexity.

The partition function $\bar{Z}$ of the spin-glass 3D Ising model in a fixed replica can be expressed as follows [3–5]:

$$\bar{Z} = (2\sinh 2\widetilde{K})^{\frac{mnl}{2}} \cdot \text{trace}(V_3 V_2 V_1)$$

(5)

$$V_3 = \prod_{j=1}^{mnl} \exp\left\{-i\widetilde{K}''\Gamma_{2j}\left[\prod_{k=j+1}^{j+mn-1} i\Gamma_{2k-1}\Gamma_{2k}\right]\Gamma_{2j+2mn-1}\right\} = \prod_{j=1}^{mnl} \exp\{i\widetilde{K}''s'_j s'_{j+mn}\}$$

(6)

$$V_2 = \prod_{j=1}^{mnl} \exp\{-i\widetilde{K}'\Gamma_{2j}\Gamma_{2j+1}\} = \prod_{j=1}^{mnl} \exp\{i\widetilde{K}'s'_j s'_{j+1}\}$$

(7)

$$V_1 = \prod_{j=1}^{mnl} \exp\{i\widetilde{K}^* \cdot \Gamma_{2j-1}\Gamma_{2j}\} = \prod_{j=1}^{mnl} \exp\{i\widetilde{K}^* \cdot C_j\}$$

(8)

Here, $\widetilde{K}^*$ is defined by $e^{-2\widetilde{K}} \equiv \tanh\widetilde{K}^*$ [2–8]. We define the matrices $C_j$ and $s'_j$ as follows:

$$C_j = I \otimes I \otimes ... \otimes I \otimes C \otimes I \otimes ... \otimes I \quad (9)$$

and

$$s'_j = I \otimes I \otimes ... \otimes I \otimes s' \otimes I \otimes ... \otimes I \quad (10)$$

For the ferromagnetic 3D Ising model, the Clifford algebra representation plays an important role in solving analytically its exact solution [3–5]. Meanwhile, the Clifford algebra representation is also important to reveal the mathematical structures of the spin-glass 3D Ising model, in which non-local behavior, non-Gaussian, and non-commutative of operators exist also.

For the spin-glass 3D Ising model, the Clifford algebra representations for the partition function $\bar{Z}$ (Equation (5)) and the transfer matrices (Equations (6)–(8)) have the almost same formulas as those of the ferromagnetic 3D Ising model, with the following differences: (1) The present interactions $\widetilde{K}$, $\widetilde{K}'$, and $\widetilde{K}''$ are randomly distributed. (2) The partition function $\bar{Z}$ for the spin glass cannot be written in terms of $\text{trace}(V_3 V_2 V_1)^m$ (and $\sum_{i=1}^{2^{nl}} \lambda_i^m$) as in the ferromagnetic case (see, for instance, Equation (2) in [4]). (3) In the products of Equations (6)–(8), j run from 1 to mnl, but in Equations (3)–(5) of [4], j run from 1 to nl. (4) For the internal factors in the transfer matrix $V_3$ (see Equation (6)), k for the product run from $j+1$ to $j+mn-1$, while in Equation (3) of [4], k run from $j+1$ to $j+n-1$. (5) The exponential factors of $s'_j s'_{j+mn}$ shows up in $V_3$ for the spin glass model (see Equation (6)), but the exponential factors of $s'_j s'_{j+n}$ appear in Equation (A10) of [4] for the ferromagnetic one. All these differences are caused by randomness of interactions in the spin-glass 3D Ising model, which do not change a fact of existing long-range entanglement between Ising spins within a plane (mn spins). The periodic boundary condition and the largest eigenvalue principle used in Zhang–Suzuki–March procedure [4] for the ferromagnetic 3D Ising model cannot be utilized for computing analytically the spin-glass 3D Ising model.

Although the Ising model with only the nearest neighboring interactions behave to be fully locally defined in the Ising spin variable language, in the transfer matrices, the set of all allowed states contribute to partition function and free energy in a way of that all spins are entangled. In the transfer matrix $V_3$, the nonlocality shows up in the alternative Clifford algebra description, defined through auxiliary fermionic Γ-operators, which reflects the global effect of the system with the nontrivial topological structures [3–5]. This is caused by the contradiction between the 2D character of transfer matrices and the 3D arrangement of spins located on a 3D lattice. The interaction between each of the two nearest neighboring spins along the third-dimension acts as a long-range engagement via a medium of the entanglement of all the spin in a plane. The nearest neighboring interaction along the third dimension behaves as an effective long-range interaction. This is indeed a pure quantum mechanics effect, being a natural character of a 3D many-body interacting spin system. Thus, such a nonlocality exists also in the space of all the Ising spin states since the descriptions in the two different spaces (with Ising spin variable language and Clifford algebra description, respectively) are connected by a series of equalities [4]. Besides the existence of the nontrivial topological structures, the spin-glass 3D Ising model possesses the characters of nonlocality, non-planarity, randomness, frustration, and non-ergodic behavior [15,47]. □

**Definition 5.** *The absolute minimum core (AMC) model of the spin-glass 3D Ising model, $M_{AMC}^{3D}$, is defined as a spin-glass 2D Ising model interacting with its nearest neighboring plane.*

**Theorem 2.** *Any algorithms, which use any approximations and/or break the long-range spin entanglement in the AMC model, $M_{AMC}^{3D}$, cannot result in the exact solution of the spin-glass 3D Ising model $M_{SGI}^{3D}$.*

**Proof of Theorem 2.** Finding the ground state of the spin-glass Ising model can be accomplished by computing H(σ) for accounting the combinatorial complexity for all $2^N$ possible configurations. The upper bound of the computational complexity of a spin-glass 3D Ising model, $C^U(M_{SGI}^{3D})$, is $O(2^N)$ [43,47,51]. In what follows, we shall determine the lower bound of the computational complexity of a spin-glass 3D Ising model, $C_L(M_{SGI}^{3D})$. Determining the ground state properties as well as the critical behavior at glassy phase transitions in disordered spin systems could be relevant for complexity theory of the satisfiability problem, due to the intractability concentration phenomena [16,27,52]. The Clifford algebra representation reveals the basic character of mathematical structures, which plays an importation role in determining the lower bound for its computational complexity [15]. The key point is that as revealed above with the Clifford algebra representation (Theorem 1), the long-range entanglement exists between spins in the spin-glass 3D Ising model, due to the nonlinear internal factors in the transfer matrices (see Equation (6)). The randomness of interactions in the spin-glass 3D Ising model does not change this character (the nature of the 3D many-body systems) but increases the computational complexity. Thus, an AMC model, $M_{AMC}^{3D}$, exists in the spin-glass 3D Ising model [15], in which the entanglements between the spins should not be broken. Indeed, the AMC model is the basic element of the spin entanglements in $M_{SGI}^{3D}$. Any algorithms, which use any approximations and/or break our AMC model, cannot find the exact solution of the spin-glass 3D Ising model. □

**Theorem 3.** $C_L(M_{SGI}^{3D}) \geq C^U(M_{AMC}^{3D})$.

**Proof of Theorem 3.** We have to consider frustration in the spin glass systems [15]. There are two cases for the computational complexity of the core model, $C(M_{SGC}^{3D})$, for computing the spin-glass 3D Ising model: 1) In some replicas, frustration is limited to occur within a plaquette, which can be included always in two neighboring planes (i.e., the AMC model). We have $C_L(M_{SGI}^{3D}) = C_L(M_{SGC}^{3D}) = C^U(M_{AMC}^{3D})$. 2) In some replicas, frustration in the 3D case may appear on closed polygons that are higher than a plaquette, which cannot be included always in two neighboring planes. We have $C_L(M_{SGI}^{3D}) = C_L(M_{SGC}^{3D}) > C^U(M_{AMC}^{3D})$. Therefore, combining the two cases, we have the following conclusion: the lower bound for computational complexity of the spin-glass 3D Ising model, $C_L(M_{SGI}^{3D})$, is equal to or larger than the computational complexity as computed by brute force search of the AMC model, $C^U(M_{AMC}^{3D})$. That is, $C_L(M_{SGI}^{3D}) \geq C^U(M_{AMC}^{3D})$. □

**3. Mapping between the Spin-Glass 3D Ising Model and the Boolean Satisfiability**

**Definition 6.** *Let $M_B^K$ be an optimization model where the upper script fixes the parameter K, and the lower indices indicate the character of the model.*

**Definition 7.** *Let $C(M_B^K)$ be the computational complexity of the model $M_B^K$.*

**Definition 8.** *Let $C^U(M_B^K)$ be the upper bound of the computational complexity of $M_B^K$. The upper bound for a model is equal to the computational complexity as computed by brute force search.*

**Definition 9.** *Let $C_L(M_B^K)$ be the lower bound of the computational complexity of $M_B^K$.*

**Theorem 4.** *The spin-glass 3D Ising model can be mapped into the Boolean satisfiability problem for K ≥ 4. Namely, $M_{SGI}^{3D} \Leftrightarrow M_{SAT}^{K \geq 4}$ and $C(M_{SGI}^{3D}) = C(M_{SAT}^{K \geq 4})$.*

**Proof of Theorem 4.** It has been known for long that links exist between statistical physics and combinatorial optimization [38,47]. The statistical physics of frustrated spin models serves to acquire a better understanding of complexity, by mapping the study of the ground states of disordered models onto the optimization problems [38,47]. In fact, the spin-glass problem is at the core of the statistical physics of disordered systems, which also deals with Boolean variables (spins), interacting with random exchange couplings [38,47]. Each pair of interacting spins can be treated as a constraint and finding the state of minimal energy in a spin-glass is equalized to minimizing the number of violated constraints. Although the precise form of the constraints in a spin glass model and a K-SAT problem somehow differ, deep similarities exist [39,53]. In both cases, the difficulty for computations comes from the existence of frustration [38,47], nonlocality [4,15,32,34], and randomness, which forbids us to find the global optimal state by a purely local optimization procedure. In what follows, we illustrate the mapping between the spin-glass 3D Ising model and the Boolean satisfiability problem.

We first focus on the K-SAT problem, which is defined as follows [35–37,39,40]. We consider N Boolean variables $\{x_i = 0; 1\}_{i=1,\ldots,N}$ and choose randomly K among $N$ possible indices i. For each of them, choose a literal that is the corresponding $x_i$ or its negation $\bar{x}_i$ with equal probabilities of one half. A clause $C$ is the logical OR of the K literals previously chosen, and $C$ will be true (or satisfied) if and only if at least one literal is true. We repeat this process to obtain M independently chosen clauses $\{C_l\}_{l=1,\ldots,M}$ and ask for all of them to be true at the same time (i.e., taking the logical AND of the M clauses). Therefore, we reach a Boolean expression in the conjunctive normal form [40], written as:

$$F = \bigwedge_{l=1}^{M} C_l = \bigwedge_{l=1}^{M} \left( \bigvee_{i}^{K} z_i^{(l)} \right) \tag{11}$$

where ∧ and ∨ stand for the logical AND and OR operations, respectively. We realize a solution of the K-SAT problem when a logical assignment of the $\{x_i\}$s satisfying all clauses, i.e., evaluating $F$ to be true. If no such assignment exists, $F$ will be unsatisfiable. For large instances ($M, N \to \infty$), mathematical analysis and numerical simulations indicate evidently that when $\alpha = M/N$ crosses a critical value $\alpha_c(K)$, the probability of finding a logical assignment of the $\{x_i\}$s satisfying all the clauses falls abruptly from one down to zero [37].

For K-SAT problems, all interactions involve K spins, and the energy of an interaction node $a$ involving spins $\sigma_{i_1}, \cdots, \sigma_{i_p}$ is represented by [36,39]:

$$E_a = 2 \prod_{r=1}^{K} \frac{(1 + J_a^r \sigma_{i_r})}{2} \tag{12}$$

It depends on a set of K coupling constants $J_a = (J_a^1, \ldots, J_a^K)$, which take values $\pm 1$. This interaction node possesses a simple interpretation as a clause: the energy $E_a$ is zero as soon as at least one of the spins $\sigma_{i_r}$ is opposite to the corresponding coupling $J_a^K$. If all spins equal their couplings, the energy is equal to 2.

We then focus on the Ising spin model: Consider a set of N Ising spins $\sigma_i \in \{\pm 1\}$ and M groups of interacting variables, which are called function nodes. A set of $n_a$ spins is involved in each function node *a*. The set of all these spins is denoted by $V_a$. The interaction is an arbitrary function of the spins in $V_a$, depending on the problem one considers and can also involve hidden variables. When we adopt the Ising spin notion, a true Boolean variable is mapped onto $S_i = +1$, whereas a false variable gives $S_i = -1$ (in Ising models with $S = 1/2$, we have $S_i = \pm 1/2$). A logical assignment *{S}* is a set of N spins $S_i$ out of all $2^N$ possible configurations. Denote the (random) set of clauses by *{C}*. We then choose the energy-cost function *H[{C}; {S}]* to be the number of clauses violated by the configuration *{S}*. The total energy of a configuration $\sigma_1, \ldots, \sigma_N$ is represented as [39]:

$$E = \sum_{a=1}^{M} E_a \tag{13}$$

and the goal in combinatorial optimization is to search a configuration of spins, which minimizes *E*. If the ground state energy is zero (respectively strictly positive), the logical clauses are satisfiable (respectively unsatisfiable). The free-energy density f of the resulting spin system at a formal temperature *T* is given by the logarithm of the partition function [35]

$$Z[\{C\}] = \sum_{\{S\}} exp\left(-\frac{H[\{C\},\{S\}]}{k_B T}\right)$$

with assumption of being self-averaging as the size $N$ of the instance of the K-SAT problem goes to infinity. In order to calculate the disorder average, one can use the replica trick:

$$\overline{lnZ} = \lim_{n\to 0} \partial_n \overline{Z^n} \tag{5}$$

where at first a positive integer number n is considered, and the replica limit n → 0 is achieved by some kind of analytical continuation in n. This problem is generalized by introducing an additional parameter $β = 1/(k_B T)$, an inverse temperature in the physics language, and studying the Boltzmann probability distribution [39]:

$$P(\sigma_1, \dots \sigma_N) = \frac{1}{Z} exp(-\beta E) \tag{6}$$

where $Z$ is a normalization constant. This arises naturally from the point of view of physics and connects directly with problems in statistical mechanics. <O> is denoted as the expectation of an observable $O$ (can be any function of the spin σ$_i$) with respect to this measure. These expressions are just those defined for the spin-glass Ising model we studied in quantum statistical mechanics. In the statistical mechanics, the temperature is an important parameter for measuring physical properties.

For spin glasses systems (described by the Edwards–Anderson model [44]), usually, all the interactions involve two spins, so all $n_a$ are equal to 2; the energy of an interaction node $a$ involving spins σ$_i$ and σ$_j$ is given by [39] $E_a = -J_{ij}\sigma_i\sigma_j$, where $J_{ij}$ is called the exchange coupling constant. The total energy of a configuration σ$_1$, …, σ$_N$ is described by Equation (13), while the partition function and the Boltzmann probability distribution are given by Equations (15) and (16), respectively. These expressions are just consistent with the Hamiltonian (1) and the partition function (2) or (5) for the spin-glass 3D Ising model if we consider random interactions.

From the first glimpse, it seems that the spin-glass Ising models correspond to the K-SAT problem with K = 2. However, as we have known, the dimensionality of the spin-glass Ising models contributes greatly to the computational complexity of the systems. The previous work [32–34] revealed that the spin-glass 2D Ising model is a P-problem, whereas the spin-glass 3D Ising model is a NP-complete problem. The 2D Ising model is a self-dual model, which is dual to a 2D Ising model with $E_a = -J_{ij}s_i s_j$, with Ising spins $s_i$ and $s_j$. However, the 3D Ising model is dual to a 3D $Z_2$ lattice gauge model with $E_a = -J_{ijpq}s_i s_j s_p s_q$, with Ising spins $s_i$, $s_j$, $s_p$, and $s_q$ around primitive squares (or plaquettes) of the lattice. Figure 1 shows the duality between the 3D $Z_2$ lattice gauge model and the original 3D Ising model [13,54,55], while Figure 2 illustrates the mapping between a two-spin link interaction in the 3D Ising model and a four-spin plaquette interaction in the 3D $Z_2$ lattice gauge model. The 3D $Z_2$ lattice gauge lattice is displaced from the original 3D Ising lattice by half a lattice spacing in each crystallographic direction [13,54–57]. The vertices of the 3D $Z_2$ lattice gauge lattice lie in the centers of the elementary cubes of the original lattice and vice versa. As shown in Equation (2.30) of ref. [55], the duality is valid as the condition $k_{\nu\lambda,i} = \frac{1}{2}(1 - r_i r_{i+\hat{\mu}})$ is held. From Equations (2.27)–(2.29) in [55], besides a factor of $(sinh 2\tilde{\beta})^{3N/2}$, the product of $C_k(\tilde{\beta})$ is the only factor associated with interactions in the partition function. As the interactions become random, the duality condition is not altered.

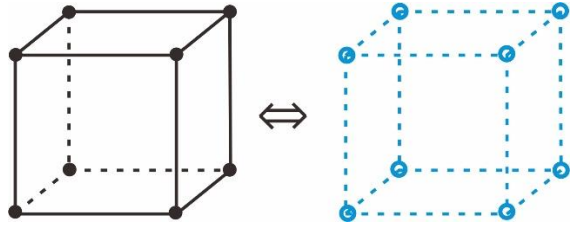

**Figure 1.** Duality between the 3D Ising model and the 3D $Z_2$ lattice gauge model.

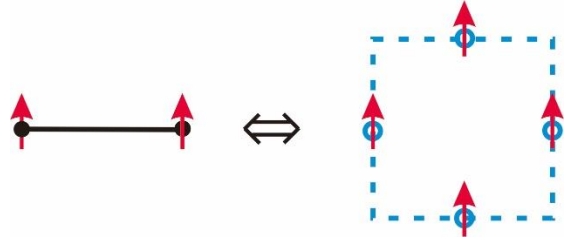

**Figure 2.** Mapping between a two-spin interaction for a link in the 3D Ising model and a four-spin interaction for a plaquette in the 3D $Z_2$ lattice gauge model.

Notice that in Figure 2, we just illustrate the cases for parallel spin alignments, but the spins can align with randomly distributed directions (even with frustrations) in the spin-glass systems. As indicated above, the randomness of interactions and spin alignments does not affect the mapping, as long as the particular constraint could be satisfied for such a mapping [13,54,55]. It clearly shows that the spin-glass 3D Ising model corresponds to the K-SAT problem with K = 4. However, in some replicas, frustration in the 3D case could appear on closed polygons, which are higher than a plaquette. Such closed polygons cannot be included always in two neighboring planes. If we consider all the possible frustrations in the 3D lattice, more than two neighboring planes must be considered, and the interactions between more than four spins need to be taken into account also. Therefore, the spin-glass 3D Ising model is equivalent to the K-SAT problem for K ≥ 4. Namely, $M_{SGI}^{3D} \Leftrightarrow M_{SAT}^{K\geq 4}$ and thus $C(M_{SGI}^{3D}) = C(M_{SAT}^{K\geq 4})$. □

## 4. Mapping between the AMC Model in the Spin-Glass 3D Ising Model and the Boolean Satisfiability

**Theorem 5.** *The AMC model in the spin-glass 3D Ising model can be mapped into the Boolean satisfiability problem for K = 3. Namely, $M_{AMC}^{3D} \Leftrightarrow M_{SAT}^{K=3}$ and $C(M_{AMC}^{3D}) = C(M_{SAT}^{K=3})$.*

**Proof of Theorem 5.** The AMC model in the spin-glass 3D Ising model is constructed by a spin-glass 2D Ising model interacting with its nearest neighboring plane [15]. We have shown that the lower bound of the computational complexity of the spin-glass 3D Ising model is that of the AMC model as computed by brute force search. That is, $C_L(M_{SGI}^{3D}) = C^U(M_{AMC}^{3D})$. We shall figure out the equivalence between the AMC model and the K-SAT problem. The AMC model $M_{AMC}^{3D}$ can be treated as a plane lattice where three links (as a star) imposed on each site (see Figure 3 for equivalence between three interactions in the AMC model and those in a star lattice) if we neglect the nontrivial topological structure of the 3D lattice. This simplification does not affect the proof of the present theorem because the nontrivial topological structure would cause even more computational complexity. Such a star model is the duality of a triangular lattice by the well-known star-triangular relation (the Yang–Baxter equation in the continuous limit) [7]. The star-triangular relation was developed firstly in electric networks [58] and is represented for Ising models as follows:

$$K_1 K_1^* = K_2 K_2^* = K_3 K_3^* = K_1 K_2 + K_2 K_3 + K_3 K_1 = \frac{K_1^* K_2^* K_3^*}{(K_1^* + K_2^* + K_3^*)}$$

(17)

where $K_i$ and $K_i^*$ ($i = 1,2,3$) are interactions for the star and triangular lattices, respectively. The Kramers–Wannier relation between the dual lattices is identified [2,54,55,59]:

$$K^* = -\frac{1}{2}ln(tanhK) \tag{7}$$

Figure 3 also represents the duality between a star lattice (i.e., the AMC model) and a triangular lattice.

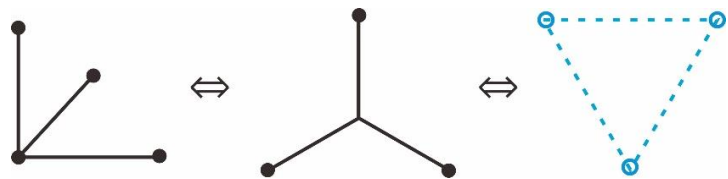

**Figure 3.** Equivalence between three interactions in the AMC model and those in a star lattice and the duality to a triangular lattice.

Then, following the procedure in [13,54,55], we can perform a similar process to dual a star lattice (i.e., the AMC model) to a 2D triangular lattice model with $E_a = -J_{ijp}s_is_js_p$, with Ising spins $s_i$, $s_j$, and $s_p$ located around primitive triangles of the lattice. Figure 4 illustrates mapping between a two-spin interaction for a link in a star lattice (i.e., the AMC model) and a three-spin interaction for a triangle in a triangular lattice. Note that in Figure 4, we just illustrate the parallel spin alignments, but the spins can align with randomly distributed directions in the spin-glass systems, even with frustrations. However, according to the discussion above, the factors related with interactions in the partition function are not associated with the condition for the mapping. Thus, we can obtain a conclusion that the randomness of interactions and spin alignments and even frustrations will not affect the mapping, as the particular constraint should be satisfied for such a mapping [13,54,55]. Therefore, the AMC model of the spin-glass 3D Ising model is equivalent to the K-SAT problem for K = 3. Namely, $M_{AMC}^{3D} \Leftrightarrow M_{SAT}^{K=3}$ and thus $C(M_{AMC}^{3D}) = C(M_{SAT}^{K=3})$. □

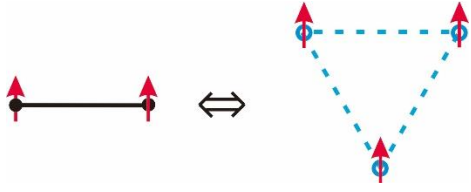

**Figure 4.** Mapping between a two-spin interaction for a link in a star lattice (i.e., the AMC model) and a three-spin interaction for a triangle in a triangular lattice.

We obtain the following theorems:

**Theorem 6**. *The lower bound of the computational complexity of the Boolean satisfiability problem for K ≥ 4 equals to the computational complexity by brute force search of the AMC model in the spin-glass 3D Ising model, which is equivalent to the computational complexity by brute force search of the Boolean satisfiability problem for K = 3. Namely, we have* $C_L(M_{SAT}^{K\geq 4}) \geq C^U(M_{AMC}^{3D}) = C^U(M_{SAT}^{K=3})$.

**Proof of Theorem 6.** It is a direct consequence of Theorems 1–5. According to Theorems 1–3, any algorithms, which use any approximations and/or break the long-range spin entanglement in the AMC model, $M_{AMC}^{3D}$, cannot result in the exact solution of the spin-glass 3D Ising model $M_{SGI}^{3D}$. $C_L(M_{SGI}^{3D}) \geq C^U(M_{AMC}^{3D})$. According to Theorem 4, $M_{SGI}^{3D} \Leftrightarrow M_{SAT}^{K\geq 4}$ and $C(M_{SGI}^{3D}) = C(M_{SAT}^{K\geq 4})$. According to Theorem 5, we have $M_{AMC}^{3D} \Leftrightarrow M_{SAT}^{K=3}$ and $C(M_{AMC}^{3D}) = C(M_{SAT}^{K=3})$ Therefore, any algorithms, which use any approximations and/or break the long-range spin entanglement in the K-SAT problem for K = 3, $M_{SAT}^{K=3}$, cannot result in the exact solution of the K-SAT problem for K ≥ 4, $M_{SAT}^{K\geq 4}$. We have $C_L(M_{SAT}^{K\geq 4}) = C_L(M_{SGI}^{3D}) \geq C^U(M_{AMC}^{3D}) = C^U(M_{SAT}^{K=3})$. □

**Theorem 7**. $M_{AMC}^{3D}$ *is the border between* $M_{SGI}^{3D}$ *and* $M_{SGI}^{2D}$, *while* $M_{SAT}^{K=3}$ *is the border between* $M_{SAT}^{K\geq 4}$ *and* $M_{SAT}^{K=2}$. †

**Proof of Theorem 7.** It is evidently correct. According to Theorems in [16,32–34], $M_{SGI}^{3D}$ and $M_{SAT}^{K\geq 4}$ are catalogued to a kind of models for NP-complete. According to Theorems 5 above, $C^U(M_{AMC}^{3D}) = C^U(M_{SAT}^{K=3})$ is the lower bound of their computational complexity. $M_{AMC}^{3D}$ is also NP-complete, but cannot be reduced furthermore to be a P-problem. By contrast, $M_{SGI}^{2D}$ and $M_{SAT}^{K=2}$ are catalogued to a kind of models for P-problem. Thus, the theorem is validated. □

† Note added that a NP-Intermediate (NPI) area exists between the NP-complete problems and the P-problems, while the AMC model is located at the border between the NPI and the NP-complete problems [60].

**Theorem 8.** *The lower bound of the computational complexity of the Boolean satisfiability problem for K ≥ 4, $C_L(M_{SAT}^{K\geq 4})$, or $C^U(M_{SAT}^{K=3})$, is in subexponential and superpolynomial.*

**Proof of Theorem 8.** It is valid as an immediate consequence of Theorem 3 of [15] and Theorem 6. According to Theorem 3 in [15], the computational complexity of the AMC model of a spin-glass 3D Ising model, $C(M_{AMC}^{3D})$, cannot be reduced to be less than $O(2^{mn})$ by any algorithms. It means that the AMC model must be computed by brute force search in order to obtain the solution of the spin-glass 3D Ising model. $O(2^{mn})$ equals to $O((1 + \varepsilon)^N)$, with $\varepsilon \to 0$ and $\varepsilon \neq 1/N$, which is subexponential, and superpolynomial [15]. According to Theorem 6, we have $C_L(M_{SAT}^{K\geq 4}) = C_L(M_{SGI}^{3D}) \geq C^U(M_{AMC}^{3D}) = C^U(M_{SAT}^{K=3})$. Thus, $C_L(M_{SAT}^{K\geq 4})$, or $C^U(M_{SAT}^{K=3})$, is in subexponential and superpolynomial. □

## 5. Conclusions

In conclusion, we have proven that the spin-glass 3D Ising model can be mapped to the K-SAT problem for K ≥ 4, that is, $M_{SGI}^{3D} \Leftrightarrow M_{SAT}^{K \geq 4}$, by the duality between the spin-glass 3D Ising model and the spin-glass 3D $Z_2$ gauge lattice theory and the consideration of random interactions and frustration. Furthermore, we have proven that the AMC model of the spin-glass 3D Ising model is equivalent to the K-SAT problem for K = 3, namely, $M_{AMC}^{3D} \Leftrightarrow M_{SAT}^{K=3}$. We have proven that $C_L(M_{SAT}^{K \geq 4}) = C_L(M_{SGI}^{3D}) \geq C^U(M_{AMC}^{3D}) = C^U(M_{SAT}^{K=3})$. $M_{AMC}^{3D}$ is the border between $M_{SGI}^{3D}$ and $M_{SGI}^{2D}$, while $M_{SAT}^{K=3}$ is the border between $M_{SAT}^{K \geq 4}$ and $M_{SAT}^{K=2}$. The lower bound of the computational complexity of the Boolean satisfiability problem for K ≥ 4, $C_L(M_{SAT}^{K \geq 4})$, is in subexponential and superpolynomial. The computational complexity of the K-SAT problem for K ≥ 4 cannot be reduced to that of the K-SAT problem for K < 3. The present work provides a bridge between mathematics, computer science, and physics, which enhances understanding and efficiency of solutions of related problems.

**Funding:** This research was funded by the National Natural Science Foundation of China under grant number 52031014.

**Institutional Review Board Statement:** Not applicable.

**Informed Consent Statement:** Not applicable.

**Data Availability Statement:** The data are available on reasonable request from the corresponding author.

**Conflicts of Interest:** The authors declare no conflicts of interest.